# Scaling population cycles of herbivores and carnivores

Christian Mulder[1,*] and A. Jan Hendriks[2]

[1] National Institute for Public Health and the Environment, Box 1, 3720BA Bilthoven, The Netherlands

[2] Department of Environmental Science, Radboud University Nijmegen, Box 9010, 6500GL, Nijmegen, The Netherlands

**ABSTRACT**

**Summary:** Periodicity in population dynamics is a fundamental issue. In addition to current species-specific analyses, allometry facilitates understanding of limit cycles amongst different species. So far, body-size regressions have been derived for the oscillation period of the population densities of warm-blooded species, in particular herbivores. Here, we extend the allometric analysis to other clades, allowing for a comparison between the obtained slopes and intercepts. The oscillation periods were derived from databases and original studies to cover a broad range of conditions and species. Then, values were related to specific body size by regression analysis. For different groups of herbivorous species, the oscillation period increased as a function of individual mass as a power law with exponents of 0.11-0.27. The intercepts of the resulting linear regressions indicated that cycle times for equally-sized species increased from homeotherms up to invertebrates. Overall, cycle times for predators did not scale to body size. Implications for these differences were addressed in the light of intra– and interspecific delays.

**Contact:** christian.mulder@rivm.nl, A.J.Hendriks@science.ru.nl

**Supplementary information:** Supplementary literature is provided as appendix. Additional models are described in arXiv:0910.5057v1.

## 1 INTRODUCTION

Scaling already fascinated Greek philosophers like Aristotle (384–322 BC). Depending upon the assumption that mind could elucidate all the laws of the universe, in *Analytica Posteriora* he saw as first that knowledge generates from the discovery of causal relationships. Still, it took other two millennia before scientists became enabled by Descartes for plotting their data on Cartesian coordinates. Ecologists were fascinated by the remarkable regularity in the population oscillations observed in laboratory and field studies. Thus, the periodicity in population dynamics is still one fundamental issue in ecology. However, the relationships between oscillation period ($\tau_o$) and body mass ($m$) have received so far surprisingly little attention. Time parameters, such as age at maturity or death, seem to scale to adult mass $m$ (e.g., Peters 1983, West *et al.* 1997, Gillooly *et al.* 2002, Brown *et al.* 2004, Hendriks and Mulder 2008), but allometric scaling for other groups, in particular hetero-therms, invertebrates and carnivores, has to be tested. Moreover, the focus has mainly been on slopes of linear (log-log) regressions while a comparison of the differences between intercepts is equally important for understanding ongoing mechanisms. In our study, we aim to check allometric scaling of periodicity in population density for a wide range of species and trophic levels. Our null-hypothesis is that the oscillation period $\tau_o$ scales to body mass $m$ with a slope of ¼, as observed for many other time variables in biology, in particular the age at maturity ($\tau_m$). As an alternative, the oscillation period $\tau_o$ may be size-independent indicating that other factors, like environmental conditions, are more important in cycle times.

## 2 METHODS

The oscillation periods were collected from laboratory experiments and field surveys reported in literature (complete references provided in the Supplementary Information as appendix). All time series, including short periods, were taken into account to obtain sufficient data for regression analyses of various species groups. Adult species' body-mass values were taken from the original studies; if adult $m$ could not be obtained directly, it was estimated from the body size, i.e. the length, of a closely related taxon. Oscillation periods $\tau_o$ were usually reported as the time between two similar phases of a cycle. Together, 759 oscillation periods covering 251 species were collected; most of those values (683 cycle times) were retrieved from one data collection with standardized entries for periodic populations, the NERC – CPB 'Global Population Dynamics Database' available online at http://www3.imperial.ac.uk/cpb/research/patternsandprocesses/gpdd. The entire data collection was further subdivided in data sets with comparable phylogenetic and ecological characteristics. Although less comparable data were available for protists, based on a few data for very dissimilar unicellulars, those regressions were derived for completeness. Following our hypotheses, oscillation periods $\tau_o$ were compared to age at maturity $\tau_m$.

## 3 RESULTS

Data on aquatic herbi-detritivores representing laboratory experiments on consumer-resource dynamics supported a close correlation between cycle times and body mass. Overall, within the common weight interval of $10^{-4} - 10^{-2}$ kg, cycle times of invertebrate herbivores increased in the sequence of aquatic, herb-dwelling and tree-dwelling species. The intercepts of the homeothermic grazers were at the lower end of the range observed for the heterotherms, about a factor of three lower (Fig. 1).

---

[*] To whom correspondence should be addressed.

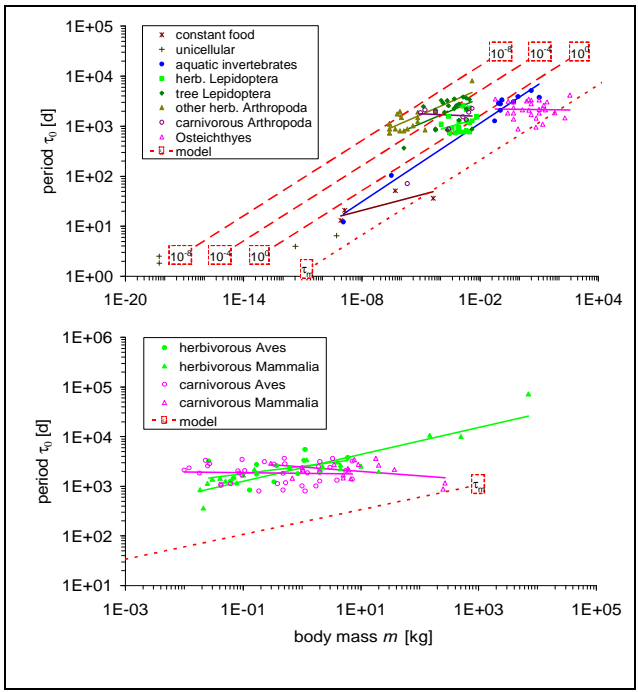

**Fig. 1.** Oscillation periods $\tau_o$ versus body mass $m$ for cold-blooded species (upper panel) and warm-blooded species (lower panel). Linear regressions (solid lines) obtained from metadata on herbivores (filled symbols) and carnivores (open symbols) were highly significant ($P < 0.0001$) only in the following cases: aquatic herbi-detritivore invertebrates ($\tau_o = 3.8 \cdot 10^3$ $(2.9 \cdot 10^3 – 5.0 \cdot 10^3) \cdot m^{0.26\ (0.23–0.30)}$, $R^2 = 97\%$), terrestrial herbivorous insects ($\tau_o = 1.2 \cdot 10^4$ $(4.2 \cdot 10^3 – 3.7 \cdot 10^4) \cdot m^{0.17\ (0.09–0.26)}$, $R^2 = 53\%$ besides Lepidoptera) and herbivorous mammals ($\tau_o = 2.4 \cdot 10^3$ $(1.9 \cdot 10^3 – 3.0 \cdot 10^3) \cdot m^{0.27\ (0.21–0.34)}$, $R^2 = 81\%$). Geometric averages given with 95% confidence interval; all the data sources are available as separate references in the enhancement.

Cycle times for carnivores were size-independent and their averages were within a factor of 1.3. Subdivision of fish data into smaller taxonomic groups did not yield size scaling either, while the recorded number of invertebrate carnivores was too low to allow subcategories. The oscillation periods for seals, the largest mammalian predators included in the data set, were close to their maturity age. Due to these data, cycle time of mammalian carnivores decreased slightly with body size.

Slopes up to about ¼ were in agreement with the range of 0.26-0.29 reported in previous studies on herbivorous homeotherms (e.g., Krukonis and Schaffer 1991, Damuth 2007). The intercept for aquatic herbi-detritivores was just over 4 times that of the age at maturity $\tau_m$, whereas the levels are more than 6 times higher for the other groups. It indicates that intraspecific delays cannot be excluded as a cause for oscillations in systems, whereas the terrestrial cycles are more likely to be (re)generated by complex trophic interactions (Murdoch *et al*. 2002, Turchin 2003). The intercepts for warm-blooded herbivores were small in comparison to those of their invertebrate equivalents. Such a difference is explainable from the high consumption and production rates in homeotherms, decreasing oscillation periods within predator–prey systems. In the case of terrestrial invertebrates, in fact, one would expect the oscillation periods to increase with plant size (Yodzis and Innes 1992), as can be concluded from a comparison of intercepts for herb- and tree-dwelling species. Indirect support for interspecific causes comes from the observation that cycle times increase with plant longevity (Hogstedt *et al*. 2005), although cycle times of insects were also related to latitude and humidity (Kendall *et al*. 1999).

In contrast to a weak relationship for predatory mammals reported in literature (Krukonis and Schaffer 1991), we found cycle times for all predators to be size-independent (Fig. 1). Thus, rather than by predator–prey interactions, oscillations for carnivores might be induced by other causes, for instance, seasonal changes in environmental conditions including effects thereof on food availability and quality. Species may override such allometric trends by synchronisation to annual cycles, as recently noted for reproduction and social behaviour. Predator's body mass may be less than linearly proportional to prey's mass at the end of the body-size spectrum, as in the case of social hunters (Carbone *et al*. 1999) or for soil-dwelling microbivores (Mulder and Elser 2009).

Finally, herbivores and specialist carnivores usually have a type-II functional response (Jeschke *et al*. 2004), generally more likely to induce limits cycles than the type-III response, with general predators that can easily shift to other resources. For instance, the occurrence of ungrazed compartments in detrital food webs (such as algae and roots) might dampen oscillations at low trophic levels in contrast to higher trophic levels. The strength of the empirical relationships between size scaling and oscillation period varies and more data are necessary for some species groups, like predatory arthropods. Still, all these consistent differences between herbivores and carnivores are remarkable and, to our knowledge, novel. We show here that macroecology provides valuable additional viewpoints to the dynamic interpretation of periodic populations.

# arXiv – APPENDIX
# Supplementary Literature

Values for some laboratory populations refer to experiments in which species were fed intermittently or continuously (Pratt 1943; Nicholson 1954; Huffaker 1958; Beddington and May 1975; Halbach 1979; Halbach *et al*. 1983). Other studies in which no additional food was added and consumers interacted with their resources (Tsuchiya *et al*. 1972; Levin *et al*. 1977; Veillleux 1979; Bohannan and Lenski 1997; Yoshida *et al*. 2003) were included as well. Data on field populations were taken from the statistical analysis of extensive time series on temperate species (Kendall *et al*. 1998). Additional information on invertebrate taxa not covered by the previous series was obtained from original studies (Utida 1957; Clark 1963; Baltensweiler 1971; Berryman 1995; Grover *et al*. 2000) and vertebrates (Newsome 1969; Southern 1970; Caughley 1976; Itô 1980; Calder 1983, 1984; Peterson *et al*. 1984; Ginzburg and Inchausti 1997; Scheffer *et al*. 1997; Turchin and Hanski 1997; Boonstra *et al*. 1998; Lambin *et al*. 2000). Together 759 oscillation periods were collected, 30 obtained in laboratory experiments and 729 derived from field surveys. Of the latter, 36 cycle times were taken from literature as originally reported and 693 values were retrieved from one current database with standardized entries for periodic populations (Kendall *et al*. 1998). Together our data collection represents 251 species. Data sets were then subdivided in groups with comparable phylogenetic and ecological characteristics. In addition, averages per individual group were calculated for comparison of oscillation characteristics that did not scale to body mass. Experiments in which food was added were assigned to a separate group. Although few comparable data were available for unicellulars, their regressions were derived for completeness ($R^2 = 91\%$, $P \equiv 0.05$). Following our hypotheses, oscillation periods $\tau_o$ were compared to maturity ages $\tau_m$.